# Theoretical and experimental study of attenuation in cancellous bone

Weiya Xie[a, †], Wenyi Xu[a, †], Dong Yu[a], Haohan Sun[a], Ying Gu[a], Xingliang Tao[a], Liming Cheng[b], Hao Wang[a,*], Qian Cheng[a, c,*]

[a]*Institute of Acoustics, School of Physics Science and Engineering, Tongji University, Shanghai, PR China*
[b]*The Key Laboratory of Spine and Spinal Cord Injury Repair and Regeneration of Ministry of Education, Tongji University, Shanghai, PR China*
[c]*Shanghai Research Institute for Intelligent Autonomous Systems, Shanghai, PR China*
[†]*These authors contributed equally to this paper.*

# Abstract

Photoacoustic (PA) technology can provide information on both the physical structure and chemical composition of bone, showing great potential in bone assessment. However, due to the complex composition and porous structure of cancellous bone, the PA signals generated and propagated in cancellous bone are complex and difficult to be directly used in cancellous bone analysis. In this paper, a photoacoustic differential attenuation spectrum (PA-DAS) method is proposed. By eliminating the PA spectrum of the optical absorption sources, the propagation attenuation characteristics of cancellous bone are studied theoretically and experimentally. An analytical solution for the propagation attenuation of broadband ultrasound waves in cancellous bone is given by applying high-frequency and viscous corrections to Biot's theory. An experimental system of PA-DAS with an eccentric excitation differential detection system is established to obtain the PA-DAS of cancellous bone and its acoustic propagation characteristic on the rabbit osteoporosis model. The PA-DAS quantization parameter-*slope* is further extracted to quantify the attenuation of high and low frequency components. The results show that the PA-DAS can distinguish osteoporotic bone from normal bone, enabling quantitative assessment of bone mineral density and the diagnosis of osteoporosis.

## 1   Introduction[1]

Osteoporosis and fractures stemming from it have emerged as chronic diseases adversely affecting the health of the elderly[1]. As populations age, the annual global increase in the number of fracture patients is approximately 8.9 million, leading to a considerable rise in public health costs[2-4]. Therefore, early diagnosis of osteoporosis can not only prevent fractures but also substantially reduce healthcare expenditures and resource usage.

---

[*] Corresponding authors: wanghao@tongji.edu.cn, q.cheng@tongji.edu.cn.

[1] Abbreviations: area ratio of trabecular bone (ARTB); bone mineral density (BMD); continuous wavelet transform (CWT); dual energy X-ray absorptiometry (DEXA); ethylenediaminetetraacetic acid (EDTA); mean trabecular thickness (MTT); photoacoustic (PA); photoacoustic time-frequency spectrum (PA-TFS); photoacoustic time-frequency spectral analysis (PA-TFSA); power-weighted mean frequency (PWMF); quantitative ultrasound (QUS); region of interest (ROI); ultrasound (US).

Compared to dense bone, the microstructure and chemical composition of cancellous bone is more sensitive to osteoporosis, as evidenced by reduced trabecular thickness, connectivity, and number, along with increased lipid content[5-8], suggesting that cancellous bone is ideally suited as a diagnostic site for early osteoporosis. However, the unique structural characteristics of cancellous bone make its detection and quantitative evaluation challenging, thereby complicating the early diagnosis of osteoporosis. The primary detection methods employed in clinical and research settings for osteoporosis include dual-energy X-ray absorptiometry (DEXA), quantitative computed tomography (QCT), magnetic resonance imaging (MRI), and quantitative ultrasound (QUS)[9-12]. DEXA serves as the "gold standard", primarily because it provides the best predictor of osteoporotic fractures through bone mineral density (BMD) information[13]. However, DEXA accounts for only 60%–70% of changes in bone strength and lacks information on microstructure and elasticity[14]. QCT extends bone analysis from two to three dimensions, providing both volumetric BMD and microstructural characteristics[15,16]. However, its utility is hampered by radiation hazards and a lack of chemical information. MRI can identify alterations in bone marrow fat content and microstructure, facilitating early diagnosis[17-19]. Nonetheless, its high cost and operational complexity limit its widespread use[20,21]. QUS, being radiation-free, quick, affordable, and user-friendly, has gained traction as a powerful tool for screening bone quality[5,22,23]. Initially centered on cortical bone[24-26], QUS is increasingly being used to explore cancellous bone characteristics due to its heightened sensitivity to osteoporotic changes. It provides physical information on BMD, bone microstructure, and mechanical properties, mainly by detecting the speed of sound (SoS) and broadband ultrasound attenuation[12,27]. Unfortunately, it falls short in detecting changes in the organic chemical composition of bone tissue.

Photoacoustic (PA) techniques offer both chemical and physical insights into biological tissues and have been utilized for tissue identification and disease detection, including osteoporosis[28-32]. Over the past decade, significant advancements have been made in PA-based bone evaluation. Mandelis and Lashkari established a set of photoacoustic-ultrasound (PA-US) backscattering detection systems, successfully detecting minute changes in the BMD of both cancellous and cortical bones[33]. Their work indicated that the apparent integral backscattering coefficient decreases with a decrease in collagen content[34-36]. Wang employed three-dimensional PA imaging (PAI) and power spectrum analysis to achieve both qualitative and quantitative evaluations of bone microstructure[37,38]. He further introduced multispectral PA decoupling and thermal PA methods for quantitative evaluation of the organic and inorganic chemical components in cancellous bone[37-39]. Additionally, a combined PA-US system was developed, verifying the in vivo feasibility of assessing human calcaneus microstructure and chemical composition[40]. Steinberg developed a multispectral PA-US dual-mode system capable of in vivo detection of fat and blood ratios in tibial bone marrow[41]. The SoS for the first arriving wave in the tibia showed a strong correlation with the SoS value based on QUS[42]. Feng and Cheng have simulated various aspects of PA in skeletal tissues, including light attenuation and distribution, along with PA signal generation and propagation, based on a three-dimensional model[43]. They proposed a PA physicochemical spectrum method for assessing changes in the chemical composition and microstructure of cancellous bone[44,45]. Our group has leveraged the MWPA time-frequency spectral analysis method to evaluate both the chemical content (minerals and lipids) and microstructure of bone tissues, based on the distinct optical absorption characteristics and sizes of various chromophores present[46,47]. The preceding research in this field demonstrates the successful application of various PA imaging and spectral analysis methods for osteoporosis diagnosis, making it possible to assess BMD, bone microstructure, and chemical composition in a comprehensive manner[48].

Nevertheless, the intricacy of PA signals, which arise from the two-phase, solid-liquid porous structure of cancellous bone, presents a significant challenge in using PA techniques for bone assessment[49]. Firstly, PA signals generated from various chemical clusters within cancellous bone with different microscales have complex optical and ultrasonic spectral distributions. In addition, these chemical clusters have uneven spatial distributions, adding another layer of complexity to the spatial distribution of PA sources. Thus, cancellous bone's PA sources possess a threefold complexity. Secondly, as these PA signals navigate through the cancellous bone, they undergo multiple scattering and attenuation events. Therefore, the final PA signal received by the transducers is a composite of broadband signals from distributed PA sources and the acoustic propagation characteristics specific to the porous cancellous bone. Decoupling these signals could yield valuable multi-dimensional insights into cancellous bone.

In this paper, we propose a PA differential attenuation spectrum (PA-DAS) method designed to remove the contribution of PA sources on PA signals in the frequency domain. This allows for a focused study—both theoretical and experimental—on the acoustic propagation characteristics of cancellous bone for bone quality assessment. Theoretically, we apply high-frequency and viscous corrections to Biot's theory, tailoring it for two-phase, solid-liquid porous media, and employ numerical simulations to validate. Subsequently, ex vivo experiments are performed to measure PA-DAS, and a quantitative PA-DAS parameter is extracted for evaluating BMD and diagnosing osteoporosis. Our results highlight the potential utility of this method for comprehensive bone quality assessment.

## 2  Modified Biot's theory

Cancellous bone is a complex porous medium composed of solid trabecular bone interspersed with fluid-filled bone marrow clusters, making it a typical example of a porous, elastic, viscous medium. Biot and Willis established the elastic theory for understanding acoustic propagation in such saturated solid-liquid two-phase porous media, thereby enabling further theoretical research in the field[50–55]. Biot's theory has subsequently found applications in the nondestructive evaluation of bone via biomedical ultrasound (US)[56–61]. However, Biot's theory is limited to cases where the acoustic frequency is below the medium's critical frequency $f_c$, and the flow of liquid through the pores is well-described by Poiseuille flow. The critical frequency $f_c$ is defined as[62]:

$$f_c = \frac{\phi \eta}{2\pi \rho_f \mathcal{K}} \qquad (1)$$

where $\phi$ denotes the porosity, $\eta$ denotes the fluid viscosity, $\rho_f$ represents the fluid density, and $\mathcal{K}$ denotes the permeability. For cancellous bone, filled with viscous bone marrow, $f_c$ typically ranges between 1-10 kHz[63,64]. The sizes of the trabecular bone vary from 50 to 200 μm, while the trabecular spaces (bone marrow clusters) range from 0.2 to 3 mm[65]. The SoS in trabecular bone and bone marrow are 3200 m/s and 1500 m/s, respectively[66]. Notably, the frequencies of PA signals generated in these structures exceed 220 kHz[67], far surpassing $f_c$. This means that the laminar flow condition described by Poiseuille's law no longer holds, necessitating modifications to Biot's theory for high-frequency applications.

In addition, the viscous nature of the fluid bone marrow leads to energy dissipation due to the relative motion between the fluid and the solid trabecular framework, further indicating the need for viscosity corrections. To account for the dissipation of acoustic wave propagation in a solid-liquid two-phase porous medium, the governing equations of motion can be expressed as follows[54,55]:

$$\mu_b \nabla^2 \vec{u} + (\lambda_b + \mu_b)\nabla e + Q\nabla \epsilon = \frac{\partial^2}{\partial t^2}(\rho_{11}\vec{u} + \rho_{12}\vec{U}) + b\frac{\partial}{\partial t}(\vec{u} - \vec{U}) \qquad (2)$$

$$Q\nabla e + R\nabla \epsilon = \frac{\partial^2}{\partial t^2}(\rho_{12}\vec{u} + \rho_{22}\vec{U}) - b\frac{\partial}{\partial t}(\vec{u} - \vec{U}). \qquad (3)$$

where $\vec{u}$ denotes the solid skeleton displacement vector, $\vec{U}$ represents the liquid displacement vector. The elastic constants $\mu_b$ and $\lambda_b$ characterize the frame's material elasticity and also depend on structural parameters such as porosity $\phi$. $R$ and $Q$ are additional elastic constants, with $Q$ being Biot's constant that describes the coupling between the liquid and solid phases. $e = e_{11} + e_{22} + e_{33} = \nabla \cdot \vec{u}$ denotes the normal strains for the solid, $\epsilon = \varepsilon_{11} + \varepsilon_{22} + \varepsilon_{33} = \nabla \cdot \vec{U}$ denotes the strain for fluid, while $\rho_{mn}$ are mass coefficients related to the porosity $\phi$ and the mass densities $\rho_s$ and $\rho_f$ of the solid and fluid $\rho_{12}$ is the mass coupling. The parameter $b = \frac{\eta \phi^2}{\mathcal{K}}$ serves as a dissipation factor and is a function of the porosity $\phi$.

The Fourier transform solution of velocities for the shear wave ($c_T^*$) and longitude wave ($c_{L1}^*$ and $c_{L2}^*$) in solid-liquid two-phase porous media with viscous losses can be expressed as follows[68]:

$$c_T^{*2} = N\left(\rho_{22} + \frac{b}{i\omega}\right) \Big/ \left[\left(\rho_{11} + \frac{b}{i\omega}\right)\left(\rho_{22} + \frac{b}{i\omega}\right) - \left(\rho_{12} - \frac{b}{i\omega}\right)^2\right]$$

$$= N\rho_{22}(\omega)/[\rho_{11}(\omega)\rho_{22}(\omega) - \rho_{12}^2(\omega)]. \qquad (4)$$

$$c_{Lj}^{*2} = \frac{\left(M + \frac{b}{i\omega}H\right) \pm \sqrt{\left(M + \frac{b}{i\omega}H\right)^2 - 4L\left(\rho_{11}\rho_{22} - \rho_{12}^2 + \frac{b}{i\omega}\rho\right)}}{2\left(\rho_{11}\rho_{22} - \rho_{12}^2 + \frac{b}{i\omega}\rho\right)}, (j = 1,2) \qquad (5)$$

where elastic coefficient $N = \mu_b$, $\rho_{11}(\omega) = \rho_{11} + \frac{b}{i\omega}$, $\rho_{12}(\omega) = \rho_{12} - \frac{b}{i\omega}$, $\rho_{22}(\omega) = \rho_{22} + \frac{b}{i\omega}$, $H = (P + R + 2Q)$, $L = (PR - Q^2)$, $M = (R\rho_{11} + P\rho_{22} - 2Q\rho_{12})$, $\rho = \rho_{11} + \rho_{22} + 2\rho_{12}$, $\omega$ denotes angular frequency.

The velocity of shear wave ($c_T$) and two types of longitude wave ($c_{L1}$ and $c_{L2}$) in solid-liquid two-phase porous media without viscosity is given by[54,55]

$$c_T^2 = N\rho_{22}/(\rho_{11}\rho_{22} - \rho_{12}^2), \tag{6}$$

$$c_{Lj}^2 = \frac{M \pm \sqrt{M^2 - 4L(\rho_{11}\rho_{22} - \rho_{12}^2)}}{2(\rho_{11}\rho_{22} - \rho_{12}^2)}, \quad (j = 1,2) \tag{7}$$

By comparing the velocities in solid-liquid two-phase porous media with and without viscosity, we observe that the viscosity loss alters Biot's mass coefficients $\rho_{mn}$ into complex quantities. This results in the shear wave velocity, as well as the fast and slow longitudinal wave velocities, becoming complex numbers. As sound waves propagate through a dissipative solid-liquid two-phase porous medium, their amplitudes decay progressively with increased propagation distance.

To account for high-frequency dissipation, the Biot's mass coefficients can be transformed as follows:

$$\rho_{mn}(\omega) = \rho_{mn} + (-1)^{m+n}\frac{bF(\kappa)}{i\omega}, \quad (m,n = 1,2). \tag{8}$$

Where, $F(\kappa) = \frac{1}{3}\frac{i^{1/2}\kappa\tanh(i^{1/2}\kappa)}{1-\frac{1}{i^{1/2}\kappa}\tanh(i^{1/2}\kappa)}$ is a complex number that represents the deviation from Poiseuille's law as frequency increases, reflecting the phase difference between velocity and friction forces[54]. The frequency-dependent dissipation coefficient $\kappa = a(\frac{2\pi f \rho_f}{\eta})^{\frac{1}{2}}$, where $a$ is the average pore size in the porous medium.

The attenuation coefficients are determined by the imaginary parts of the complex wave numbers:

$$\alpha_T^* = -\text{Im}(k) = -\text{Im}(\frac{2\pi f}{c_T^*}), \tag{9}$$

$$\alpha_{Lj}^* = -\text{Im}(k) = -\text{Im}(\frac{2\pi f}{c_{Lj}^*}), \quad (j = 1,2). \tag{10}$$

The high-frequency dissipation coefficient $\alpha^*$ is influenced by several factors, including the porosity $\phi$, the average pore size $a$ of the solid-liquid two-phase porous medium, the frequency $f$ of the acoustic wave, and the fluid viscosity $\eta$. Therefore, the porosity $\phi$ of cancellous bone can be inferred based on the frequency-dependent attenuation coefficient.

## 3 Numerical simulations

*3.1 Numerical simulation parameters*

Based on the above modified Biot's theory, we numerically simulated the propagation of acoustic waves in porous cancellous bone using MATLAB software. The parameters used for these simulations are detailed in Tables 1 and 2. It should be noted that, in general, the porosity of normal cancellous bone is about 0.73[69,70], and that of osteoporotic cancellous is larger than this value, so we mainly carried out research on the porosity range of 0.72-0.90. We specifically examined the influences of porosity and sound frequency on sound velocity and attenuation.

Table 1 Structural and acoustic parameters of cancellous bone[65].

| Parameters | Value |
| --- | --- |
| Young's modulus of solid bone $E_s$ | 22 GPa |
| Poisson's ratio of solid bone $v_s$ | 0.32 |
| Poisson's ratio of skeletal frame $v_b$ | 0.32 |
| Compressibility modulus of the solid $\kappa_s$ | 20.37 GPa |
| Solid density $\rho_s$ | 1960 kg/m³ |
| Fluid density $\rho_f$ | 1000 kg/m³ |
| Compressibility modulus of the solid $\kappa_f$ | 2.28 GPa |
| Fluid viscosity $\eta_f$ | 0.001 Pa·s |
| Power index $n$ | $n(\theta) = 1.43\cos^2(\theta) + 2.14\sin^2(\theta)$ |
| Tortuosity $\alpha_\infty$ | $\alpha_\infty = 1 - 0.259\left(1 - \frac{1}{\phi}\right) + 0.864\cos^2(\theta)$ |
| Pore size $a$ | 1 mm |

Table 2 Permeability and pore size of different porosity[71,72].

| Porosity $\phi$ | Permeability $k_0$ $(m^2)$ |
| --- | --- |
| 0.72 | 5 × 10⁻⁹ |
| 0.75 | 7 × 10⁻⁹ |
| 0.80 | 2 × 10⁻⁹ |
| 0.83 | 3 × 10⁻⁹ |
| 0.90 | 8 × 10⁻⁹ |

*3.2 Numerical simulation results*

Based on the modified Biot's theory above, we establish a computational model of semi-infinite cancellous bone and numerically simulate the propagation of PA waves in cancellous bone generated by bone marrow on the bone surface irradiated by a pulsed spot light with a diameter of 2 mm. Figure 1(a) and 1(b) demonstrate the PA field with the frequency of 2.3 MHz in the cancellous bone with 0.72 porosity and 0.83 porosity, respectively, at 17.92 μs, including fast longitude wave (fast P-wave), slow longitude wave (slow P-wave), shear wave, and Rayleigh wave. The numerical simulation shows that the speed and amplitude of fast P-waves are much larger than those of other wave modes and other wave modes, which means that we can extract very clean fast P-waves from the PA signal in the time domain.

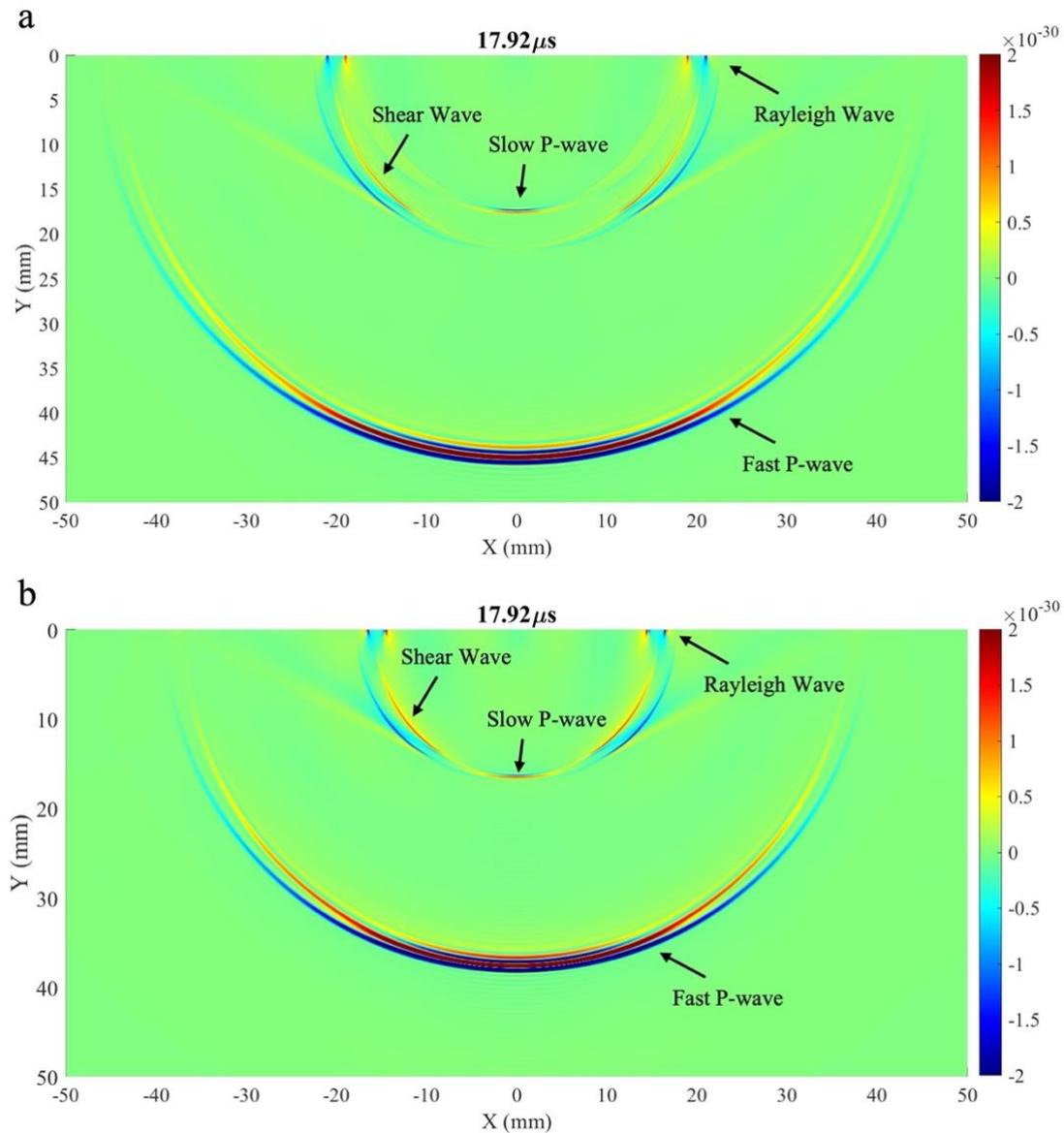

**Fig. 1** Numerical simulation results of the sound field of 2.3 MHz acoustic wave propagating in cancellous bone with (a) 0.72 porosity and (b) 0.83 porosity at 17.92 μs.

Figure 2(a) and (b) illustrate the calculated trend of the velocity of fast and slow P-waves propagating in solid-liquid two-phase porous media with porosity and frequency, respectively. Similarly, in the calculated porosity range of 0.72 to 0.90, the velocity of fast P-wave ($c_f$) is much high than that of slow P-wave ($c_s$) at same porosity. Notably, both fast and slow P-wave velocities are insensitive to frequency change, that is, there is almost no dispersion, which is very advantageous for PA detection because the bandwidth of the PA signal is usually wide. However, both velocities decrease significantly with increasing porosity, which is also consistent with Fig. 1. This is mainly due to the fact that as the porosity increases, the proportion of the solid phase decreases and the solid-liquid coupling becomes weaker, leading to a decrease in the velocity of the fast and slow waves, respectively.

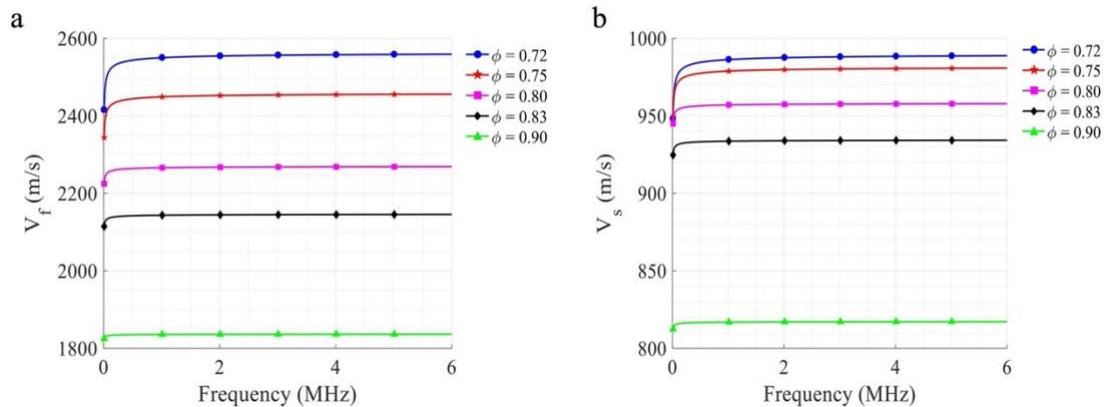

Fig. 2 Numerical simulation results. Speed of sound of (a) fast and (b) slow longitude waves with different porosity and frequency.

Figure 3(a) and (b) showcase the viscous absorption attenuation coefficients for fast and slow P-waves as they travel through solid-liquid two-phase porous media, respectively. In the calculated porosity range of 0.72 ~ 0.90, the absorption attenuation coefficient for the fast P-wave is much smaller than that of the slow P-wave with the same porosity. In addition, these attenuation coefficients are influenced by both the porosity of the cancellous bone and the frequency of the acoustic waves. When porosity is constant, the absorption attenuation coefficients for both fast and slow waves increase with increasing frequency. The attenuation coefficients exhibit a shift from fast to slow changes, eventually tending toward a linear pattern within the frequency range of 1 to 6 MHz as the frequency increases. Conversely, when frequency is held constant, the absorption attenuation coefficient decreases as porosity increases. This is mainly due to the fact that the higher the porosity of cancellous bone, the smaller the solid-liquid interface area, which in turn leads to a reduction in energy dissipation caused by the relative motion between the solid bone trabecular frame and the liquid bone marrow.

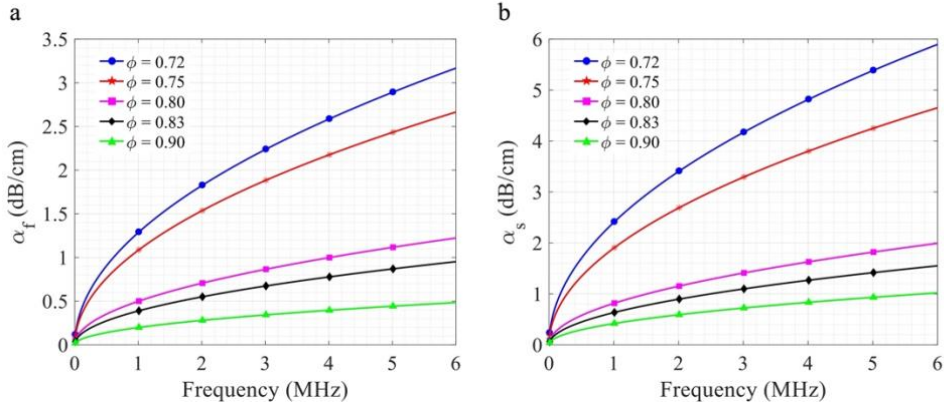

**Fig. 3** Numerical simulation results. Viscous absorption attenuation coefficients of (a) fast and (b) slow longitude waves with different porosity and frequency.

These simulation results indicate that the fast P-wave with high speed and low attenuation is more suitable for the PA detection of cancellous bone, and the porosity $\phi$ of porous media can be evaluated based on the attenuation of acoustic waves across various frequencies.

## 4 Photoacoustic differential attenuation spectrum method

As discussed earlier, the PA signal arriving at the transducers is a complex mixture of broadband signals originating from distributed PA sources and the inherent acoustic propagation characteristics of porous cancellous bone. The PA signal received by the transducers at time $t$ can be represented as:

$$p(t) = p(t_0) * h(t - t_0) \tag{11}$$

where $*$ denotes convolution operator, $p(t_0)$ represents the PA signal generated by bone tissue received by the transducers at time $t_0$, $h(t)$ denotes the systematic response of bone tissue. Applying the Fourier transform to Eq. (11) yields the spectrum of PA signals arriving at the transducers at time $t$:

$$P(\omega) = P_0(\omega) H(\omega) e^{j\omega t_0} \tag{12}$$

where $P(\omega)$, $P_0(\omega)$, and $H(\omega)$ denote the Fourier transforms of $p(t)$, $p(t_0)$, and $h(t)$, respectively. By setting $t_0$ to 0, we can derive the frequency-related acoustic propagation characteristics of cancellous bone using the following Eq. (13) to calculate the differential attenuation spectrum of the PA signals received by transducers at different times:

$$H(\omega) = \frac{P_1(\omega)}{P_0(\omega)} \tag{13}$$

According to Fig. 3, it is evident that the attenuation coefficient is approximately linear within the 1–6 MHz range. Thus, the acoustic propagation characteristics within this frequency band can be linearly fitted to quantify attenuation across different frequencies.

## 5 *Ex vivo* experiments on rabbit bone specimens

Based on the results of theoretical and numerical simulations, the PA experiment was conducted on a rabbit osteoporosis model to verify the feasibility of bone evaluation based on PA-DAS.

*5.1 Animal model and bone tissue sample*

In our study, we used a sample of eleven 5-month-old female New Zealand white rabbits. Six were randomly selected to undergo ovariectomy, forming the experimental group, while the remaining five received a sham operation to serve as the control group. After five months of identical living conditions, all rabbits were euthanized. The left metaphyseal region was then carefully dissected and sectioned into slices with a uniform thickness of 1.5 mm. Dense peripheral bone tissue was removed, and the slices were further trimmed to a standard width $D_b$ of 10 mm, as shown in Fig. 4(a), making them suitable for the PA experiments.

*5.2 Gold standard - BMD*

Following the PA experiments, all eleven bone samples from both the experimental and control groups were subjected to micro-computed tomography scanning (Micro-CT, venus001, Avatar3, Life Medical Technology). Three-dimensional images of both osteoporotic and normal bone are presented in Fig. 4(a). As evidenced by Fig. 4(b), the BMD of the normal bones in the control group was significantly higher than that of the osteoporotic bones in the experimental group ($p<0.01$). In Fig. 4(d), the statistical analysis of the region of interest (ROI) from Fig. 4(c) reveals a significantly higher porosity in the osteoporotic group compared to the control group ($p<0.05$). Literature suggests that higher BMD is inversely related to porosity[69,70,73,74]. Therefore, it can be concluded that osteoporotic bone exhibits higher porosity compared to normal bone.

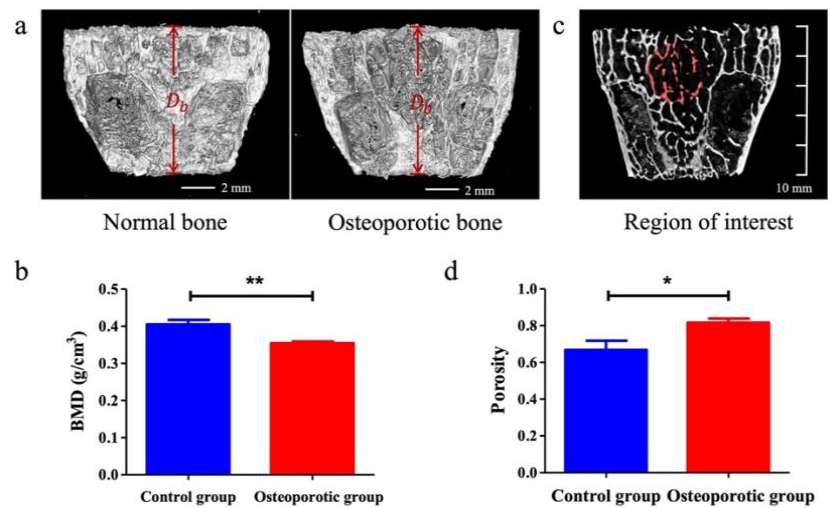

**Fig. 4** Micro-CT results. (a) 3D images of normal bone and osteoporotic bone. (b) Statistical analysis results of BMD of bone specimens from the osteoporotic group (n=6) and control group (n=5) (** $p<0.01$). (c) Region of interest to calculate porosity. (d) Statistical analysis results of porosity of bone specimens from the osteoporotic group (n=6) and control group (n=5) (* $p<0.05$).

## 5.3 PA experimental setup

To measure the frequency-dependent attenuation of PA signals in cancellous bone, we propose an eccentric excitation differential detection system specifically for PA experiments. Figure 5 illustrates the schematic layout of the experimental setup. A tunable optical parametric oscillator laser (Phocus Mobile, OPOTEK, Carlsbad, CA) produces laser pulses with durations ranging from 2 to 5 ns. We selected a laser wavelength of 1730 nm, which corresponds to the specific absorption wavelength of lipids—a major component of bone marrow clusters—to irradiate the bone samples and thereby excite PA signals[75,76]. To compensate for variations in laser energy over time, a spectrophotometer with a 9:1 transmittance-to-reflectance ratio was used to split the laser beam into two paths. One path was focused using a convex lens to irradiate a blackbody, while the other was weakly focused on one side of the bone tissue sample surface, tangent to the side of the sample, forming a light spot with a diameter $D_l$ of approximately 2 mm. The bone sample was placed on a 5 cm thick phantom to mitigate any direct light or sound reflections from the platform. As shown in Fig. 5, a needle hydrophone T1 (HNC1500, ONDA Corp., Sunnyvale, CA) with a bandwidth of 0-20 MHz was placed on the side of the sample near the light spot to receive the unattenuated PA signals. These signals were then amplified by 25 dB using an amplifier (5072PR, Olympus Corp, Tokyo, Japan) and subjected to 1 MHz high-pass filtering to remove low-frequency noise. On the opposite side of the light source, a planar transducer with a center frequency of 2.25 MHz (Olympus Corp, Tokyo, Japan) was positioned to receive the PA signals transmitted through the cancellous bone. These signals were amplified by another 25 dB using an amplifier (5073PR, Olympus Corp, Tokyo, Japan). Both transducers were acoustically coupled to the bone sample using a transparent ultrasonic coupling agent. Concurrently, a 1 MHz focusing transducer (Olympus Corp, Tokyo, Japan) was employed to receive the PA signal generated by the blackbody. Data acquisition was performed using a digital oscilloscope (HDO6000, Teledyne Lecroy, USA) set at a sampling rate of 250 MHz, which was deemed sufficient for our experimental requirements. Due to the anisotropic properties of cancellous bone, we employed a translation stage to move the bone sample longitudinally in 2 mm increments, for a total of three. This enabled us to capture PA signals at four distinct positions, thus providing a comprehensive evaluation of signal attenuation throughout the cancellous bone. To improve the signal-to-noise ratio, PA signals were recorded 50 times at each position and subsequently averaged.

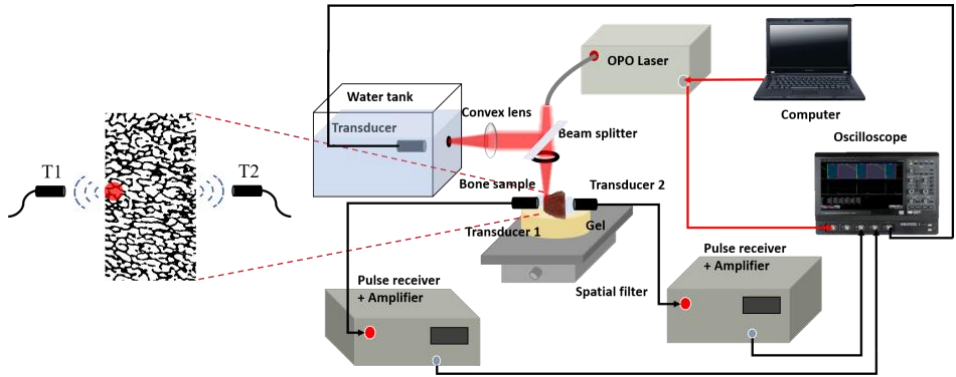

**Fig. 5** Schematic of experimental setup for PA measurement of bone samples.

## 5.4 PA signal processing

As shown in Fig.5, the PA signal received by hydrophone T1 closest to the light source serves as the unattenuated signal generated by the cancellous bone. Conversely, the PA signal received by transducer T2 located farther from the light source is a composite, reflecting both the inherent properties of the PA source and the acoustic propagation medium. By calculating the PA-DAS for both transducers, we can isolate and understand the frequency-related propagation characteristics specific to cancellous bone.

In the experiment, only longitudinal wave signals can be picked up because the gel is used as coupling agent. Moreover, the numerical results in Section 3.2 show that the attenuation of the slow P-wave is much greater than that of the fast P-wave, and the speed is much smaller than that of the fast P-wave. Therefore, it is more meaningful to analyze the fast P-wave signals with low attenuation than the complete signal.

It is necessary to select a reasonable time window for more accurate spectral analysis, so we introduce the SoS for bone marrow $c_m = 1500\ m/s$ which is larger than slow P-wave and smaller than fast P-wave, and bone trabecular $c_t = 3200\ m/s^{66}$ which is larger than fast P-wave, as the rising and falling edges, respectively, of the window for picking up fast-wave signals. As delineated in the red dotted box in Fig. 6(a), the direct PA signal duration for T1 is given by $t1 = \frac{D_l}{c_m} = 1.33$ μs. The earliest and latest times for the PA signal of T2 were selected by $\frac{D_b - D_l}{c_t}$, and $\frac{D_b}{c_m}$, respectively. The length of the PA signal as captured by T2 is $t2 = \frac{D_b}{c_m} - \frac{D_b - D_l}{c_t} = 4.17$ μs.

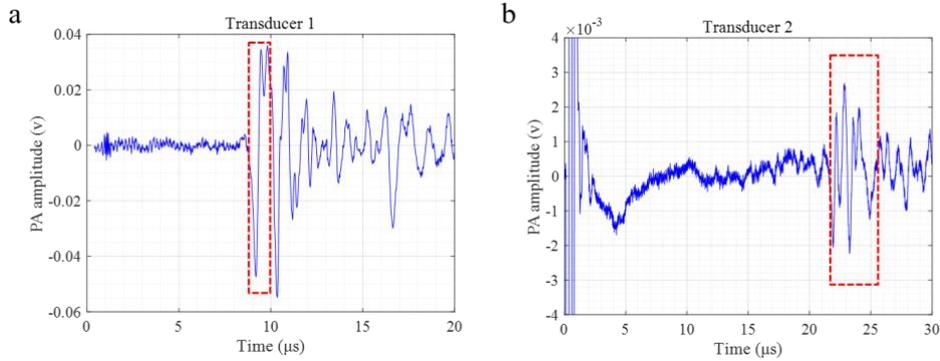

**Fig. 6** PA signals generated by bones received by (a) the transducer near the light source T1 and (b) the transducer away from the light source T2.

The frequency-response curve of the transducer T2, centered at 2.25 MHz, is presented in Fig. 7(a). As evident in this curve, the transducer produces its maximum output at the central frequency of 2.25 MHz and attenuates on either side, leading to distortion in the frequency domain of the PA signal as displayed by the oscilloscope. To more accurately investigate the spectral characteristics of the PA signal, it is necessary to correct the observed output signals in line with the frequency-response curve of the transducer. Corresponding to a 20 dB decline from the maximum frequency response to prevent excessive correction of high-frequency noise, a fitting frequency band ranging from 0.85 to 3.6MHz was selected. After applying a 1 MHz high-pass filter to eliminate the influence of low-frequency noise, the spectral characteristics of the PA signal received by the distant transducer—both before and after frequency-response correction—are illustrated by the blue and red curves in Fig. 7(b). Post-correction, the PA signal spectrum exhibited a 20 dB decrease after 1 MHz (the valid starting point after high-pass filtering) at 3.6 MHz, warranting the choice of 3.6 MHz as the upper-frequency limit. While the frequency-response curve of the broadband hydrophone T1 is basically uniform, the signals received by T1 do not need correction. Given the data in Fig. 3, the linear fitting for the frequency curve was confined to the 1 to 3.6 MHz range.

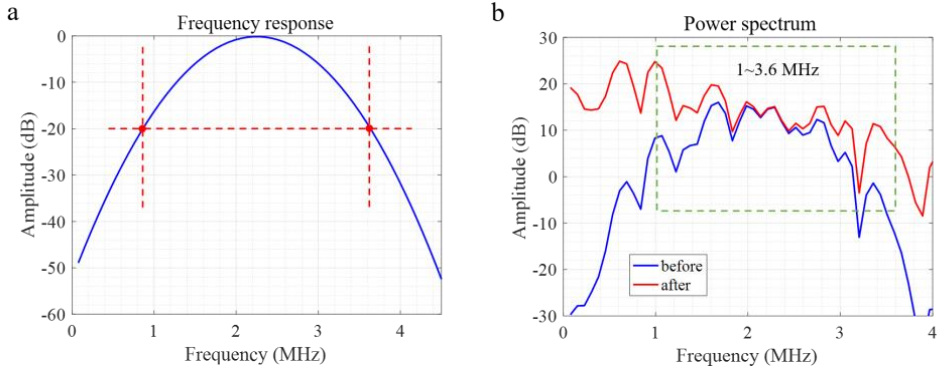

**Fig. 7** (a) Frequency-response curve of transducer. Red intersection point is frequency corresponding to maximum value decreasing by 20 dB. (b) Spectrum of bone photoacoustic signal received by transducer (solid blue line) and spectrum after frequency-response correction (solid red line).

To account for the spectral differences of the PA source caused by the anisotropy of cancellous bone, we calculated the PA-DAS $As(f)$ was calculated, based on Eq. (14) to evaluate the acoustic propagation characteristics of the bone.

$$As(f) = 10\lg\left(\frac{PSD_2(f)}{PSD_1(f)}\right) \qquad (14)$$

Here, $PSD_1$ and $PSD_2$ represent the power spectra of PA signals received by Transducers 1 and 2, respectively, as depicted in Fig. 8(a). Notably, the PA signal from the transducer closer to the light source includes a higher concentration of high-frequency components due to the absence of propagation attenuation. In contrast, the PA signal from the transducer situated farther from the light source is mainly composed of low-frequency components, owing to significant propagation attenuation in cancellous bone.

Linear fitting was performed for the PA-DAS in the frequency range of 1 to 3.6 MHz. From this, we extracted the *slope* as a quantization parameter, enabling us to obtain the frequency-related attenuation of the PA signal as it travels through the bone.

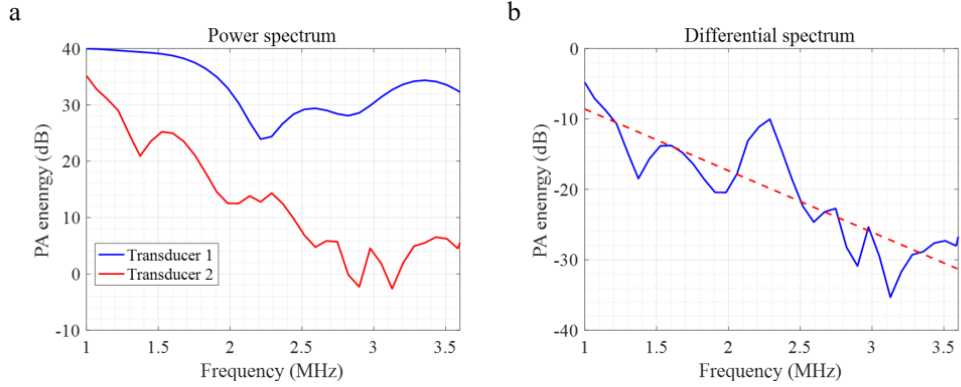

**Fig. 8** (a) Spectrum of PA signals received by the two transducers. (b) Differential spectrum of PA signal received by the two transducers.

*5.5 Results*

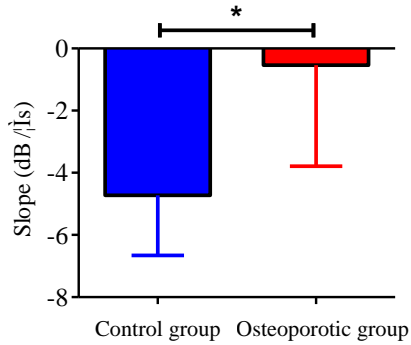

**Fig. 9** Statistical results of bone *slope* in two groups (* $p < 0.05$).

Figure 9 presents the *slopes* for both normal and osteoporotic bones, revealing a significantly larger *slope* for osteoporotic bones when compared to normal ones ($p<0.05$). Given that normal bone has higher BMD and lower porosity, the attenuation of high-frequency components is stronger during PA signal propagation through the cancellous bone structure. This results in a decreased high-to-low frequency ratio in the spectral distribution, which in turn leads to a smaller *slope* for the control group. Conversely, osteoporotic bones have a larger *slope*, which is due to the weakening of the solid-liquid interface coupling and the decrease of viscosity attenuation caused by lower BMD and higher porosity.

## 6 Conclusion and discussion

The propagation of acoustic waves through the cancellous bone, a solid-liquid two-phase porous medium was studied theoretically, numerically, and experimentally in this paper. Biot's theory was modified to account for high frequencies and viscosity, providing an analytical solution for the broadband PA signal's behavior in cancellous bone. Numerical simulation results show that when porosity is greater than 0.72, the viscous absorption attenuation of PA signals increases with frequency but decreases with greater porosity. Also, the fast P-wave with high speed and low attenuation is more suitable for the detection of cancellous bone. In PA detection, to isolate the acoustic propagation properties of porous cancellous bone from the aliasing with the PA source, a PA-DAS method was employed and a parameter-*slope* was extracted to quantify frequency-specific acoustic attenuation. Experimental data on a rabbit osteoporotic model demonstrate that PA-DAS can effectively distinguish normal from osteoporotic bone, validating its utility for bone evaluation.

Cancellous bone's acoustic propagation properties are predominantly governed by absorption and scattering of viscous solid-liquid two-phase porous structure[77]. Absorption attenuation is an energy dissipation process caused by viscous friction at the fluid-solid interface, while scattering attenuation[78] arises from reflections and scattering due to acoustic disparities between solid trabecular bone and liquid bone marrow, leading to the reduction of acoustic energy along the original propagation direction[79]. This work focuses on high-frequency viscous absorption attenuation in porous media, refining Biot's theory to address higher frequency bands in PA signals and the viscosity of bone marrow. However, for more accurate quantification of scattering characteristics of cancellous bone, future studies should consider the anisotropy and heterogeneity intrinsic to its structure.

When considering the clinical application of PA-DAS, it is crucial to account for the layers of cortical bone and soft tissue that envelop cancellous bone. Since the transducers are located on opposite sides of the bone, the thickness of the dense bone and soft tissue on both sides of the bone should be considered. Our previous studies successfully isolated PA signals originating from soft tissue, cortical bone, and cancellous bone in the time domain[45]. By using the speed of sound to calculate the thickness of these three types of media, and employing the exponential attenuation law to consider the optical and ultrasonic propagation attenuation of cortical bone and soft tissue, we can compensate the PA signals in further studies to isolate and analyze the pure PA signal corresponding to cancellous bone[80-82].

## Acknowledgments

This project was supported by The National Natural Science Foundation of China [grant numbers 12034015, 11827808]; Program of Shanghai Academic Research Leader [grant number 21XD1403600]; Postdoctoral Science Foundation of China [grant number 2019M65156], and the Fundamental Research Funds for the Central Universities. We would like to thank Editage (www.editage.cn) for English language editing.

# References

1. S. R. Cummings and L. J. Melton, "Epidemiology and outcomes of osteoporotic fractures," Lancet 359(9319), 1761-1767 (2002) [https://doi.org/10.1016/S0140-6736(02)08657-9].
2. J. Y. Reginster and N. Burlet, "Osteoporosis: A still increasing prevalence," Bone 38(2) Supplement 1, S4-S9 (2006) [https://doi.org/10.1016/j.bone.2005.11.024].
3. J. J. Jolly *et al.*, "Protective effects of selected botanical agents on bone," *Int. J. Environ. Res. Public Health* **15**(5), 963 (2018) [https://doi.org/10.3390/ijerph15050963].
4. C. Cooper, G. Campion, and L. J. Melton, "Hip fractures in the elderly: A world-wide projection," *Osteoporosis Int.* **2**(6), 285-289 (1992) [https://doi.org/10.1007/BF01623184].
5. K. A. Wear, "Mechanisms of interaction of ultrasound with cancellous bone: A review," IEEE Trans. Ultrason. Ferroelectr. Freq. Control 67(3), 454-482 (2020) [https://doi.org/10.1109/TUFFC.2019.2947755].
6. E. Seeman, "Age- and menopause-related bone loss compromise cortical and trabecular microstructure," J. Gerontol. A Biol. Sci. Med. Sci. 68(10), 1218-1225 (2013) [https://doi.org/10.1093/gerona/glt071].
7. Vesterby, H. J. Gundersen, F. Melsen, "Star volume of marrow space and trabeculae of the first lumbar vertebra: Sampling efficiency and biological variation," Bone 10(1), 7-13 (1989) [https://doi.org/10.1016/8756-3282(89)90140-3].
8. J. F. Griffith, "Age-related changes in the bone marrow," Curr. Radiol. Rep. 5(6), 24 (2017) [https://doi.org/10.1007/s40134-017-0218-8].
9. P. Pisani *et al.*, "Screening and early diagnosis of osteoporosis through X-ray and ultrasound based techniques,", *World J. Radiol.* **5**(11), 398-410 (2013) [https://doi.org/10.4329/wjr.v5.i11.398].
10. J. A. Kanis, "Assessment of fracture risk and its application to screening for postmenopausal osteoporosis: Synopsis of a WHO report," *Osteoporosis Int.* **4**, 368-381 (1994) [https://doi.org/10.1007/BF01622200].
11. C. V. Albanese, F. De Terlizzi, and R. Passariello, "Quantitative ultrasound of the phalanges and DXA of the lumbar spine and proximal femur in evaluating the risk of osteoporotic vertebral fracture in postmenopausal women," *Radiol. Med.* **116**(1), 92-101 (2011) [https://doi.org/10.1007/s11547-010-0577-1].
12. C. F. Njeh, C. M. Boivin, and C. M. Langton, "The role of ultrasound in the assessment of osteoporosis: A review," *Osteoporosis Int.* **7**(1), 7-22 (1997) [https://doi.org/10.1007/BF01623454].
13. S. R. Cummings et al., "Risk factors for hip fracture in white women. Study of Osteoporotic Fractures Research Group," N. Engl. J. Med. 332(12), 767-773 (1995) [https://doi.org/10.1056/NEJM199503233321202].
14. C. C. Glüer et al., "Three quantitative ultrasound parameters reflect bone structure," Calcif. Tissue Int. 55(1), 46-52 (1994) [https://doi.org/10.1007/BF00310168].
15. J. E. Adams, "Quantitative computed tomography," Eur. J. Radiol 71(3), 415-424 (2009) [https://doi.org/10.1016/j.ejrad.2009.04.074].

16. X. S. Liu et al., "Bone density, geometry, microstructure, and stiffness: Relationships between peripheral and central skeletal sites assessed by DXA, HR-pQCT, and cQCT in premenopausal women," J. Bone Miner. Res. 25(10), 2229-2238 (2010) [https://doi.org/10.1002/jbmr.111].
17. C. S. Gee et al., "Validation of bone marrow fat quantification in the presence of trabecular bone using MRI," J. Magn. Reson. Imaging 42(2), 539-544 (2015) [https://doi.org/10.1002/jmri.24795].
18. H. Yu et al., "Multiecho reconstruction for simultaneous water-fat decomposition and T2* estimation," J. Magn. Reson. Imaging 26(4), 1153-1161 (2007) [https://doi.org/10.1002/jmri.21090].
19. H. Yu et al., "Field map estimation with a region growing scheme for iterative 3-point water-fat decomposition," Magn. Reson. Med. 54(4), 1032-1039 (2005) [https://doi.org/10.1002/mrm.20654].
20. J. F. Griffith et al., "Vertebral bone mineral density, marrow perfusion, and fat content in healthy men and men with osteoporosis: Dynamic contrast-enhanced MR imaging and MR spectroscopy," Radiology 236(3), 945-951 (2005) [https://doi.org/10.1148/radiol.2363041425].
21. J. F. Griffith et al., "Vertebral marrow fat content and diffusion and perfusion indexes in women with varying bone density: MR evaluation," *Radiology* **241**(3), 831-838 (2006) [https://doi.org/10.1148/radiol.2413051858].
22. K. Y. Chin and S. Ima-Nirwana, "Calcaneal quantitative ultrasound as a determinant of bone health status: What properties of bone does it reflect?," *Int. J. Med. Sci.* **10**(12), 1778-1783 (2013) [https://doi.org/10.7150/ijms.6765].
23. M. Matsukawa, "Bone ultrasound," *Jpn. J. Appl. Phys.* **58**(SG), SG0802 (2019) [https://doi.org/10.7567/1347-4065/ab0dfa].
24. Q. Grimal and P. Laugier, "Quantitative ultrasound assessment of cortical bone properties beyond bone mineral density," *IRBM* **40**(1), 16-24 (2019) [https://doi.org/10.1016/j.irbm.2018.10.006].
25. D. Ta *et al.*, "Measurement of the dispersion and attenuation of cylindrical ultrasonic guided waves in long bone," *Ultrasound Med. Biol.* **35**(4), 641-652 (2009) [https://doi.org/10.1016/j.ultrasmedbio.2008.10.007].
26. Q. Grimal *et al.*, "Quantitative ultrasound of cortical bone in the femoral neck predicts femur strength: Results of a pilot study," *J. Bone Miner. Res.* **28**(2), 302-312 (2013) [https://doi.org/10.1002/jbmr.1742].
27. J. J. Kaufman and T. A. Einhorn, "Perspectives: Ultrasound assessment of bone," *J. Bone Miner. Res.* **8**(5), 517-525 (1993) [https://doi.org/10.1002/jbmr.5650080502].
28. S. Huang *et al.*, "Interstitial assessment of aggressive prostate cancer by physio-chemical photoacoustics: An *ex vivo* study with intact human prostates," *Med. Phys.* **45**, 4125-4132 (2018) [https://doi.org/10.1002/mp.13061].
29. S. Mallidi *et al.*, "Prediction of tumor recurrence and therapy monitoring using ultrasound-guided photoacoustic imaging," *Theranostics* **5**(3), 289-301 (2015) [https://doi.org/10.7150/thno.10155].

30. R. Cao *et al.*, "Functional and oxygen-metabolic photoacoustic microscopy of the awake mouse brain," *NeuroImage* **150**, 77-87 (2017) [https://doi.org/10.1016/j.neuroimage.2017.01.049].
31. J. T. Oh *et al.*, "Three-dimensional imaging of skin melanoma in vivo by dual-wavelength photoacoustic microscopy," *J. Biomed. Opt.* **11**(3), 34032 (2006) [https://doi.org/10.1117/1.2210907].
32. L. Z. Xiang and F. F. Zhou, "Photoacoustic imaging application in tumor diagnosis and treatment monitoring,", *KEM* **364-366**, 1100-1104 (2007) [https://doi.org/10.4028/www.scientific.net/KEM.364-366.1100].
33. B. Lashkari and A. Mandelis, "Coregistered photoacoustic and ultrasonic signatures of early bone density variations," *J. Biomed. Opt.* **19**(3), 36015 (2014) [https://doi.org/10.1117/1.JBO.19.3.036015].
34. L. Yang *et al.*, "Photoacoustic and ultrasound imaging of cancellous bone tissue," *J. Biomed. Opt.* **20**(7), 076016 (2015) [https://doi.org/10.1117/1.JBO.20.7.076016].
35. L. Yang *et al.*, "Bone Composition Diagnostics: Photoacoustics versus ultrasound," *Int. J. Thermophys.* **36**(5-6), 862-867 (2015) [https://doi.org/10.1007/s10765-014-1701-6].
36. B. Lashkari, L. Yang, and A. Mandelis, "The application of backscattered ultrasound and photoacoustic signals for assessment of bone collagen and mineral contents," *Quant. Imaging Med. Surg.* **5**(1), 46-56 (2015) [https://doi.org/10.3978/j.issn.2223-4292.2014.11.11].
37. X. Wang *et al.*, "Photoacoustic measurement of bone health: A study for clinical feasibility" in 2016 IEEE International Ultrasonics Symposium (IUS), pp. 1-4, IEEE, Tours, France (2016) [https://doi.org/10.1109/ULTSYM.2016.7728418].
38. T. Feng *et al.*, "Characterization of bone microstructure using photoacoustic spectrum analysis" in A. A. Oraevsky, L. V. Wang, Eds., p. 93234I, San Francisco, CA (2015) [https://doi.org/10.1117/12.2078258].
39. T. Feng *et al.*, "Bone assessment via thermal photo-acoustic measurements," *Opt. Lett.* **40**(8), 1721-1724 (2015) [https://doi.org/10.1364/OL.40.001721].
40. T. Feng *et al.*, "Functional photoacoustic and ultrasonic assessment of osteoporosis: A clinical feasibility study,", *BME Front.* **2020**, 1-15 (2020) [https://doi.org/10.34133/2020/1081540].
41. Steinberg *et al.*, "First-in-human study of bone pathologies using low-cost and compact dual-wavelength photoacoustic system," *IEEE J. Select. Topics Quantum Electron.* **25**(1), 1-8 (2019) [https://doi.org/10.1109/JSTQE.2018.2866702].
42. M. Weiss *et al.*, "Reference database for bone speed of sound measurement by a novel quantitative multi-site ultrasound device," *Osteoporos. Int.* **11**(8), 688-696 (2000) [https://doi.org/10.1007/s001980070067].
43. T. Feng *et al.*, "The feasibility study of the transmission mode photoacoustic measurement of human calcaneus bone in vivo," *Photoacoustics* **23**, 100273 (2021) [https://doi.org/10.1016/j.pacs.2021.100273].
44. T. Feng *et al.*, "Characterization of multi-biomarkers for bone health assessment based on photoacoustic physicochemical analysis method," *Photoacoustics* **25**, 100320 (2022) [https://doi.org/10.1016/j.pacs.2021.100320].

45. T. Feng *et al.*, "Feasibility study for bone health assessment based on photoacoustic imaging method," *Chin. Opt. Lett.* **18**(12), 121704 (2020) [https://doi.org/10.3788/COL202018.121704].
46. W. Xie *et al.*, "Photoacoustic characterization of bone physico-chemical information," *Biomed. Opt. Express* **13**(5), 2668-2681 (2022) [https://doi.org/10.1364/BOE.457278].
47. W. Xie *et al.*, "Bone microstructure evaluation by photoacoustic time-frequency spectral analysis" in 2020 IEEE International Ultrasonics Symposium (IUS), pp. 1-4, IEEE, Las Vegas, NV (2020) [https://doi.org/10.1109/IUS46767.2020.9251672].
48. E. A. Gonzalez and M. A. L. Bell, "Photoacoustic imaging and characterization of bone in medicine: Overview, applications, and outlook," *Annu. Rev. Biomed. Eng.* **25**, 207-232 (2023) [https://doi.org/10.1146/annurev-bioeng-081622-025405].
49. Z. E. A. Fellah *et al.*, "Ultrasonic wave propagation in human cancellous bone: Application of Biot theory," *J. Acoust. Soc. Am.* **116**(1), 61-73 (2004) [https://doi.org/10.1121/1.1755239].
50. M. A. Biot, "General theory of three-dimensional consolidation," *J. Appl. Phys.* **12**(2), 155-164 (1941) [https://doi.org/10.1063/1.1712886].
51. M. A. Biot, "Theory of elasticity and consolidation for a porous anisotropic solid," *J. Appl. Phys.* **26**(2), 182-185 (1955) [https://doi.org/10.1063/1.1721956].
52. M. A. Biot, "Theory of deformation of a porous viscoelastic anisotropic solid," *J. Appl. Phys.* **27**(5), 459-467 (1956) [https://doi.org/10.1063/1.1722402].
53. M.A. Biot, and D. G. Willis, "The elastic coefficients of the theory of consolidation," *J. Appl. Mech.* **24**(4): 594-601 (1957) [https://doi.org/10.1115/1.4011606].
54. M. A. Biot, "Theory of propagation of elastic waves in a fluid-saturated porous solid. II. Higher frequency range," *J. Acoust. Soc. Am.* **28**(2), 179-191 (1956) [https://doi.org/10.1121/1.1908241].
55. M. A. Biot, "Theory of propagation of elastic waves in a fluid-saturated porous solid. I. Low-frequency range," *J. Acoust. Soc. Am.* **28**(2), 168-178 (1956) [https://doi.org/10.1121/1.1908239].
56. F. J. Fry and J. E. Barger, "Acoustical properties of the human skull," *J. Acoust. Soc. Am.* **63**(5), 1576-1590 (1978) [https://doi.org/10.1121/1.381852].
57. R. B. Ashman, J. D. Corin, and C. H. Turner, "Elastic properties of cancellous bone: Measurement by an ultrasonic technique," *J. Biomech.* **20**(10), 979-986 (1987) [https://doi.org/10.1016/0021-9290(87)90327-7].
58. Z. A. Fellah *et al.*, "Application of the Biot model to ultrasound in bone: Direct problem," *IEEE Trans. Ultrason. Ferroelectr. Freq. Control* **55**(7), 1508-1515 (2008) [https://doi.org/10.1109/TUFFC.2008.826].
59. N. Sebaa *et al.*, "Application of the Biot model to ultrasound in bone: Inverse problem," *IEEE Trans. Ultrason. Ferroelectr. Freq. Control* **55**(7), 1516-1523 (2008) [https://doi.org/10.1109/TUFFC.2008.827].
60. M. Sadouki *et al.*, "Ultrasonic propagation of reflected waves in cancellous bone: Application of Biot theory" in 2015 6th European Symposium on Ultrasonic Characterization of Bone, pp. 1-4, IEEE, Corfu, Greece (2015) [https://doi.org/10.1109/ESUCB.2015.7169900].

61. M. Pakula *et al.*, "Application of Biot's theory to ultrasonic characterization of human cancellous bones: Determination of structural, material, and mechanical properties," *J. Acoust. Soc. Am.* **123**(4), 2415-2423 (2008) [https://doi.org/10.1121/1.2839016].
62. M. R. J. Wyllie, G. H. F. Gardner, and A. R. Gregory, "Studies of elastic wave attenuation in porous media," *Geophysics* **27**(5), 569-589 (1962) [https://doi.org/10.1190/1.1439063].
63. E. R. Hughes *et al.*, "Ultrasonic propagation in cancellous bone: A new stratified model," *Ultrasound Med. Biol.* **25**(5), 811-821 (1999) [https://doi.org/10.1016/S0301-5629(99)00034-4].
64. E. R. Hughes *et al.*, "Estimation of critical and viscous frequencies for Biot theory in cancellous bone," *Ultrasonics* **41**(5), 365-368 (2003) [https://doi.org/10.1016/S0041-624X(03)00107-0].
65. Bennamane and T. Boutkedjirt, "Theoretical and experimental study of the ultrasonic attenuation in bovine cancellous bone," *Appl. Acoust.* **115**, 50-60 (2017) [https://doi.org/10.1016/j.apacoust.2016.08.011].
66. T. Cox *et al.*, "k-space propagation models for acoustically heterogeneous media: Application to biomedical photoacoustics," *J. Acoust. Soc. Am.* **121**(6), 3453-3464 (2007) [https://doi.org/10.1121/1.2717409].
67. M. W. Sigrist and F. K. Kneubühl, "Laser-generated stress waves in liquids," *J. Acoust. Soc. Am.* **64**(6), 1652-1663 (1978) [https://doi.org/10.1121/1.382132].
68. T. Feng *et al.*, "Photoacoustic bone characterization: A progress review," *Chin. Sci. Bull.* **68**(26), 3437-3454 (2023) [https://doi.org/10.1360/TB-2023-0335].
69. Hosokawa and T. Otani, "Ultrasonic wave propagation in bovine cancellous bone," *J. Acoust. Soc. Am.* **101**(1), 558-562 (1997) [https://doi.org/10.1121/1.418118].
70. M. L. McKelvie and S. B. Palmer, "The interaction of ultrasound with cancellous bone," *Phys. Med. Biol.* **36**(10), 1331-1340 (1991) [https://doi.org/10.1088/0031-9155/36/10/003].
71. H. W. Wang *et al.*, "Label-Free bond-selective imaging by listening to vibrationally excited molecules," *Phys. Rev. Lett.* **106**(23), 238106 (2011) [https://doi.org/10.1103/PhysRevLett.106.238106].
72. H. Lei *et al.*, "Characterizing intestinal inflammation and fibrosis in Crohn's disease by photoacoustic imaging: Feasibility study," *Biomed. Opt. Express* **7**(7), 2837-2848 (2016) [https://doi.org/10.1364/BOE.7.002837].
73. L. Cardoso *et al.*, "In vitro acoustic waves propagation in human and bovine cancellous bone," *J. Bone Miner. Res.* **18**(10), 1803-1812 (2003) [https://doi.org/10.1359/jbmr.2003.18.10.1803].
74. D. Porrelli *et al.*, "Trabecular bone porosity and pore size distribution in osteoporotic patients – A low field nuclear magnetic resonance and microcomputed tomography investigation," *J. Mech. Behav. Biomed. Mater.* **125**, 104933 (2022) [https://doi.org/10.1016/j.jmbbm.2021.104933].
75. B. Li *et al.*, "Characterization of a rabbit osteoporosis model induced by ovariectomy and glucocorticoid," *Acta Orthop.* 81(3), 396-401 (2010) [http://doi.org/10.3109/17453674.2010.483986].

76. A.W. Ebaerhardt, A.Y. Jones, and H. C. Blair, "Regional trabecular bone matrix degeneration and osteocyte death in femora of glucocorticoid treated rabbits," Endocrinology, 142(3), 1333-1340 (2001) [https://doi.org/10.1210/endo.142.3.8048].
77. P. Laugier, "Quantitative ultrasound of bone: Looking ahead," *Joint Bone Spine* **73**(2), 125-128 (2006) [https://doi.org/10.1016/j.jbspin.2005.10.012].
78. R. Strelitzki, P. H. F. Nicholson, and V. Paech, "A model for ultrasonic scattering in cancellous bone based on velocity fluctuations in a binary mixture," *Physiol. Meas.* **19**(2), 189-196 (1998) [https://doi.org/10.1088/0967-3334/19/2/006].
79. M. F. Insana *et al.*, "Describing small-scale structure in random media using pulse-echo ultrasound," *J. Acoust. Soc. Am.* **87**(1), 179-192 (1990) [https://doi.org/10.1121/1.399283].
80. Pifferi *et al.*, "Optical biopsy of bone tissue: A step toward the diagnosis of bone pathologies," *J. Biomed. Opt.* **9**(3), 474-480 (2004) [https://doi.org/10.1117/1.1691029].
81. Liu *et al.*, "The analysis and compensation of cortical thickness effect on ultrasonic backscatter signals in cancellous bone," *J. Appl. Phys.* **116**(12), 124903 (2014) [https://doi.org/10.1063/1.4896258].
82. M. Firbank *et al.*, "Measurement of the optical properties of the skull in the wavelength range 650-950 nm," *Phys. Med. Biol.* **38**(4), 503-510 (1993) [https://doi.org/10.1088/0031-9155/38/4/002].